\documentstyle[12pt]{article}
\date{}
\title{Dynamics with unitary phase operator:\\
 implications for Wigner's problem}
\author{{Ramandeep S. Johal}\\
{\it Institut f\"{u}r Theoretische Physik,}\\
{\it Technische Universit\"{a}t Dresden, }\\
{\it  01062 Dresden, Germany.}\\
e-mail: rjohal$@$theory.phy.tu-dresden.de\\
Ph.:  + 49 (0351) 463 35582  \\
Fax:  + 49 (0351) 463 37299}
\begin{document}
\maketitle
\def\be{\begin{equation}}
\def\ee{\end{equation}}
\def\ba{\begin{eqnarray}}
\def\ea{\end{eqnarray}}
\baselineskip 24pt
\begin{abstract}
We show that for general deformations of $SU(2)$ algebra, 
the dynamics in terms of ladder operators is preserved.
This is done for a system of precessing magnetic
dipole in magnetic field,  using the unitary phase operator
which arises in the polar decomposition
of $SU(2)$ operators.  It is pointed out that there is a
single phase operator dynamics underlying
the dynamics of usual and deformed ladder operators.   
\end{abstract}
{\bf PACS code(s)}: 03.65.Fd, 02.20.Uw, 42.50.D.  
\newpage
\section{Introduction}
Wigner's problem  \cite{1}  usually formulated for the case 
of quantum harmonic oscillator states that
the equations of motion do not determine a unique set of commutation
relations for the observables. 
In classical mechanics also, it is 
known \cite{2}-\cite{4} that same dynamical equations may be obtained using 
alternative hamiltonians and definitions of
Poisson brackets. 
Parastatistics is another example of such a nonuniqueness \cite{5}.  
Recently, Wigner's problem  for a 
precessing magnetic dipole (with dynamical algebra $SU(2)$)
  was discussed \cite{6} and a class
of modified commutation relations were shown
to be compatible with the same dynamical equations.
In this paper, we point out that
invariance of dynamics under general deformations of the  $SU(2)$  
algebra, can be understood in a unified manner as 
an underlying  dynamics in terms of unitary phase operator,
that arises in the polar decomposition of the ladder operators. 

The polar decomposition procedure referred to above is the
operator analogue of factorising a complex number into
a real argument and an exponential phase.
 For an operator the factors should
be a hermitian part and a unitary phase operator.
The unitary phase operator ($e^{i\phi}$) in turn defines
a hermitian phase operator $\phi$. 
For the  purpose of $SU(2)$ algebra, 
 the phase or  angle operator  is conjugate
to angular momentum component, 
though the canonical conjugacy is modified when, as in this 
case, the operators are bounded \cite{7,7a}. Moreover,
 dynamics in terms of unitary phase operator also
helps to understand the concept of angular velocity
in finite space quantum mechanics \cite{jo}.

In section 2, we first review the dynamics and algebraic 
structure  of the precessing
magnetic dipole in the presence of magnetic field, in terms
of generators of $SU(2)$ algebra. Then we describe
the polar decomposition procedure for the ladder operators
of this algebra and the dynamics is cast in terms of the
unitary phase operator.
It is shown that dynamics for standard ladder operators
can be derived from the dynamical equation for phase operator.
In section 3, we consider general deformations of the $SU(2)$
algebra and show that dynamics for deformed ladder operators
also follows from the same dynamical equation for phase operator.
Other classes of deformations are discussed in Section 4.
Section 5 presents some concluding remarks.
\section{Precessing magnetic dipole}
The hamiltonian for a magnetic dipole precessing in 
a magnetic field is given by $H = -\mu (\vec{J}.\vec{B})$.
For simplicity, let us choose the magnetic field to be along the $z$-axis,
so that 
\be
H = -\mu B J_z,
\label{eq1}
\ee
where $J_z$ is the $z$ component of angular momentum
operator $\vec{J}$. In terms of the ladder operators
defined by $J_{\pm}  = (J_x \pm iJ_y)/
\sqrt{2}$, and $J_0=J_z$, which are generators of 
$SU(2)$ algebra
\be
[J_0,J_{\pm}] = \pm J_{\pm},\quad [J_+,J_-] = 2J_0,
\label{e3a}
\ee
 the equations of motion are given as
\ba
{dJ_{\pm}\over dt} &=& \mp i\mu B J_{\pm},
\label{eq2a}\\
{dJ_0\over dt} &=&  0. \label{eq2b}\ea
We choose the set of basis states to be the standard angular
momentum states $\{|j,-j\rangle, |j,-j+1\rangle, \cdots,
|j,j-1\rangle,|j,j\rangle \}$, which define a $(2j+1)$-dimensional
 irreducible representation for  $J_{\pm},J_0$:
\ba
{J}_{\pm} &=& \sum_{m=-j}^{+j} \sqrt{(j\mp m)(j\pm m+1)}
|j,m\pm 1\rangle \langle jm|,\\
{J}_0 &=&  \sum_{m=-j}^{+j} m |jm\rangle \langle jm|.
\ea
The casimir for this algebra is given by 
\be
C \equiv \vec{J}^2 = J_-J_+ +J_0(J_0+1) = J_+J_- +J_0(J_0-1).
\label{e3b}
\ee
In the following we write the equations of motion 
using the unitary phase operator,   
 which arises in the polar decomposition  
 procedure  \cite{8} of ladder operators  
\ba
J_+ &=& \sqrt{J_+J_-}\; e^{i\phi} = e^{i\phi}\sqrt{J_-J_+},\label{eq8}\\ 
J_- &=& \sqrt{J_-J_+}\;e^{-i\phi} = e^{-i\phi}\sqrt{J_+J_-}.
\label{e4}
\ea
The exponential phase operator $e^{i\phi}$  is unitary
($e^{i\phi}e^{-i\phi}=e^{-i\phi} e^{i\phi}=1$) and is
given as
\be
e^{i\phi} = \sum_{m=-j}^{j-1} |j,m+1\rangle\langle j,m|
+e^{i(2j+1)\theta_0} |j,-j\rangle\langle j,j|.
\label{cr}
\ee
In other words, operator $\phi$ is hermitian.
Here $\theta_0$ is an arbitrary phase angle, which defines
the domain of phase operator $\phi$ to be $[\theta_0,\theta_0+2\pi )$.

We can write the equation of motion for $e^{i\phi}$
\be 
{d\over dt}e^{i\phi} = {1\over i\hbar}[e^{i\phi},H],
\label{eun}
\ee
using  the following commutator \cite{7}
\be
[e^{\pm i\phi},J_0] = \pm \hbar \{ -e^{\pm i\phi} +(2j+1)
                      e^{\pm i(2j+1)\theta_0}|\pm (-j)\rangle \langle \pm j|\}.
\label{comm}
\ee
Now to get  equation of motion for $J_+$, we just multiply
 Eq. (\ref{eun}) with $\sqrt{J_+J_-}$ on the left 
(or with $\sqrt{J_-J_+}$ on the right) and use the fact that
 $J_+J_-(J_-J_+)$ commutes with $J_0$ and hence with the hamiltonian.
Also we use $J_+|j,j\rangle = \langle j,j|J_- = 0$.
Similarly, we can obtain Eq. (\ref{eq2a}) corresponding to $J_-$
starting with equation of motion for $e^{-i\phi}$ and 
 using $J_-|j,-j\rangle = \langle j,-j|J_+ = 0$.

\section{Dynamics with deformed ladder operators}
In this section, we show how 
starting from the equation of motion for $e^{\pm i\phi}$, 
we can preserve the linear dynamics in terms of deformed 
ladder operators.
Specifically, we consider general deformations of $SU(2)$ algebra
\cite{11}
\be
[\tilde{J}_0,\tilde{J}_{\pm}] = \pm \tilde{J}_{\pm},
\quad [\tilde{J}_+,\tilde{J}_-] = f(\tilde{J}_0).
\label{e12}
\ee
$f(z)$ is a real parameter-dependent  analytic function of its argument,
holomorphic in the neighbourhood of zero and goes to $2z$ for certain
limiting value of the parameter.
Also define a function $g$ through
\be
f(\tilde{J}_0) = g(\tilde{J}_0) - g(\tilde{J}_0-1).
\label{e13}
\ee
 The function $g(\tilde{J}_0)$
is not unique and is determined upto any periodic function of unit period.
The casimir for this algebra is
 $\tilde{C} = \tilde{J}_-\tilde{J}_+ +g(\tilde{J}_0) = \tilde{J}_+
\tilde{J}_- +g(\tilde{J}_0-1)$.

A  generalized map which does not preserve hermitian
conjugation between $\tilde{J}_+$ and $\tilde{J}_-$ 
 may be given as
\be
\tilde{J}_+ = J_+ A(C,J_0), \quad \tilde{J}_- = B(C,J_0)J_-,
\quad \tilde{J}_0 = J_0.
\label{gena}
\ee
It is easy to verify the first relation in Eq. (\ref{e12}).
To satisfy the second relation, the following condition must hold
\be
A(J_0-1)B(J_0-1)(C-J_0(J_0-1)) -
B(J_0)A(J_0)(C-J_0(J_0+1)) = f(J_0).
\label{genm}
\ee
Assuming that $A$ and $B$ commute,  the above condition implies
\be
A(J_0-1)B(J_0-1)(C-J_0(J_0-1)) = -g(J_0-1) +p(J_0),
\ee
where $p(J_0)$ is some periodic function of period unity.
Note that this only fixes the product $A(J_0)B(J_0)$.
Different choices of these functions as well as 
the function $p$, produce a variety of realizations for 
the deformed algebra.

Now we observe  that 
\ba
~~\tilde{J}_+ = J_+A(C,J_0) &=& e^{i\phi}\sqrt{J_-J_+}\;A(C,J_0) \label{ana}\\
                          &\equiv& e^{i\phi}G(C,J_0),
\label{eqg}
\ea
where Eqs. (\ref{eq8}) and (\ref{e3b}) have been used.
Then, multiplying Eq. (\ref{eun})  on the right 
by $G(C,J_0)$ and using the fact that $G$ 
 commutes with the hamiltonian, we 
obtain the  dynamical equation for $\tilde{J}_+$
\be
{d\tilde{J_{+}}\over dt} = - i\mu B \tilde{J_{+}}.
\label{e6}
\ee 
Similarly we can write 
\ba
~~\tilde{J}_- = B(C,J_0)J_- &=& B(C,J_0)\sqrt{J_-J_+}e^{-i\phi} \label{ana2}\\
                          &\equiv&  K(C,J_0)e^{-i\phi}.
\label{eqg2}
\ea
Again multiplying the equation of motion for $e^{-i\phi}$ on the left 
by $K(C,J_0)$ and using the fact that $K$ 
 commutes with the hamiltonian, we 
obtain the  dynamical equation for $\tilde{J}_-$
\be
{d\tilde{J_{-}}\over dt} =  i\mu B \tilde{J_{-}}.
\label{e7}
\ee 
Thus we see that equations of motion for 
$\tilde{J_{\pm}}$ are identical in form 
to those for $J_{\pm}$, Eqs. (\ref{eq2a}).
This can also be proved without using
the unitary phase operator, as was done in \cite{6}.
But the idea  here 
is to point out that {\it  
underlying the  identical dynamics of the 
usual and deformed ladder operators, there is a single unitary
phase operator dynamics}. 
Note that it is not possible to obtain the dynamics
of unitary phase operator by going in the opposite
fashion, i.e. starting with the equation of motion
for ladder operators and using the polar decomposition.
This way the second term on the right hand side of 
Eq. (\ref{cr}) cannot be reproduced. 

\section{Alternate deformations}
It is clear from above that the existence of a mapping function
which commutes with the hamiltonian (for the present purpose
the mapping function is a function of operators $C$ and $J_0$)
ensures to preserve the dynamics in terms of 
 operators $J_{\pm,0}$ and $\tilde{J}_{\pm,0}$.
In this section, we study other $q$-deformations of 
the $SU(2)$ algebra which can be treated with the above scheme.

When $\tilde{J}_- =  {\tilde{J}_+}^{\dag}$ is imposed, we can  
express the deformed generators in terms of those of $SU(2)$ algebra
as the following maps 
\be
\tilde{J}_+ = \sqrt{{f(J_0+j)f(J_0-1-j)\over (J_0+j)
(J_0-1-j)}} J_+, \quad \tilde{J}_- =  {\tilde{J}_+}^{\dag},
\quad \tilde{J}_0 = J_0.
\ee
The Drinfeld-Jimbo deformation of
$SU(2)$ algebra, denoted by $SU_q(2)$ \cite{9a,9b},
is a special case of the above deformed algebras,
where $f(x)=(q^x-q^{-x})/(q-q^{-1})\equiv [x]_q$ with $q$ as real.
Its $(2j+1)$-dimensional representation is given by
\ba
\tilde{J}_{\pm} &=& \sum_{m=-j}^{+j} \sqrt{[j\mp m]_q[j\pm m+1]_q}
|j,m\pm 1\rangle \langle jm|,\\
\tilde{J}_0 = J_0 &=& \sum_{m=-j}^{+j} m |jm\rangle \langle jm|.
\ea
Note that action of $\tilde{J}_0$ is not  deformed.
The casimir of this algebra is
$\tilde{C} = \tilde{J}_-\tilde{J}_+ +[\tilde{J}_0]_q[\tilde{J}_0+1]_q
 = \tilde{J}_+
\tilde{J}_- + [\tilde{J}_0]_q [\tilde{J}_0-1]_q.$
For $q$ as phase factor, 
the above representation suffers from problem of 
negative norm. To ensure positive norm,    
a modified  representation  may be taken, 
as considered in \cite{fuji}.

Similarly, deforming maps for well known quantum algebras 
corresponding to  $SU(2)$  can be given \cite{cz}, 
and they can be discussed within this framework. 
As another example, consider Witten's second deformation 
 generated by $\{W_{\pm}, W_0\}$  and given  as follows \cite{w1}
\ba
~~[W_0,W_+]_r &\equiv & rW_0 W_+ -{1\over r}W_+W_0 = W_+, \\
~~[W_+,W_-]_{1/r^2} &=& W_0,  \\
~~[W_0,W_-]_r &=& W_-.
\ea
Here $W_{\pm,0}$ is equivalent to $\tilde{J}_{\pm,0}$.
The deforming map in terms of generators of $SU(2)$ is given by
\ba
W_0  &=& {1\over r-1/r}\left(1 - {r^{2j+1} +r^{-2j-1}\over r+1/r}r^{-2J_0}
           \right), \\
W_+  &=& r^{-J_0}\sqrt{{2r\over r+1/r}}\sqrt{[J_0+j]_r[J_0-1-j]_r
           \over (J_0+j)(J_0-1-j)} J_+, \\
W_-  &=&  {W_+}^{\dag}.
\ea
Now we find that  for $H =-\mu B J_0$,
\be
{dW_{\pm}\over dt} = -i\mu B[W_\pm,J_0] = \mp i\mu BW_{\pm},
\ee
and
\be
{dW_0\over dt} = 0, \ee
which is identical in form with Eqs. (\ref{eq2a}) and (\ref{eq2b}).

Clearly, adopting the
polar decomposition procedure for the ladder operators $J_{\pm}$ of $SU(2)$
algebra, the dynamical equations for deformed operators,
identical in form to Eqs. (\ref{e6}) and (\ref{e7}), 
can be recovered from  dynamics of the unitary phase operator.

Finally, we argue that realization of deformed ladder
operators as proposed in \cite{6} is a special case
of the mapping in Eq. (\ref{gena}).
Following \cite{6},  a non-linear deformation 
of ladder operators may be defined  through an arbitrary function
$F(C,J_0)$   as
\be
\tilde{J}_+ = {J}_+F, \tilde{J}_- = {J}_-F,   \tilde{J}_0 = {J}_0F.
\ee
The transformation leads to the following deformed algebra
\ba
~~~[\tilde{J}_0,\tilde{J}_{\pm}] &=&\left\{1-{F(C,J_0\mp 1)\over  
F(C,J_0)}\right\}
\tilde{J}_0\tilde{J}_{\pm}\pm F(C,J_0\mp 1)
\tilde{J}_{\pm}, \label{fd}\\ 
~~~[\tilde{J}_+,\tilde{J}_- ] &=& \left\{1-{F(C,J_0+ 1)\over 
F(C,J_0-1)}\right\}
\tilde{J}_+\tilde{J}_- +2F(C,J_0+1)\tilde{J}_0.
\label{fd2}
\ea
Now although the operator $\tilde{J}_0$ is modified as compared
to $J_0$ in the above algebra, the hamiltonian is  expressed in
terms of the usual operator $J_0$. Thus the commutator 
$[\tilde{J}_0,\tilde{J}_{\pm}]$ does not play any role in 
the dynamics considered in \cite{6}.
In fact, if we choose not to deform $J_0$ as considered above in 
our approach, Eq. (\ref{fd})  just reduces to 
$[{J}_0,\tilde{J}_{\pm}] =  \pm \tilde{J}_{\pm}$.
Secondly, the commutator in Eq. (\ref{fd2}) 
follows if we  choose 
$A(C,J_0) = F(C,J_0)$ and $B(C,J_0) = F(C,J_0+1)$,  
in Eqs. (\ref{gena}) and (\ref{genm}).  
 
\section{Concluding  Remarks}
We close with a few remarks on the analogous
problem for the harmonic oscillator.
It is known that a unitary phase operator does not
exist for quantum harmonic oscillator, 
 the system
for which Wigner originally formulated his problem.
A unitary phase operator was defined  for
the finite $(s+1)$-dimensional harmonic oscillator as \cite{pb}
\be
e^{i\Phi} = \sum_{n=0}^{s} |n-1\rangle\langle n|
+e^{i(s+1)\phi_0} |s\rangle\langle 0|,
\label{cro}
\ee
where $\{|n\rangle\}_{0,1,...,s}$ are the eigenstates of 
the number operator $N=\sum_{n=0}^{s}n|n\rangle\langle n|$.
 Then with the oscillator
Hamiltonian $H=N\hbar\omega$, we can write the following equation
of motion
\be
{d\over dt}e^{i\Phi} = -i\omega\{e^{i\Phi}-e^{i(s+1)\phi_0} (s+1)
 |s\rangle\langle 0|\}.
\label{dyo}
\ee
Multiplying the above equation from right with $\sqrt{N}$
and noting that annihilation operator $a = e^{i\Phi}\sqrt{N}$, we get
\be
{da\over dt} = -i\omega a.
\label{os1}
\ee
Similarly the equation of motion for creation operator $a^{\dag}$ can be
derived  
\be
{da^{\dag}\over dt} = i\omega a^{\dag}.
\label{os2}
\ee
These results match with the corresponding equations valid
for usual infinite dimensional harmonic oscillator.  
Now $q$-deformed oscillator with  
$q =e^{i2\pi/(s+1)}$ also admits finite dimensional
realizations and the above hermitian or unitary  phase operator can be
used \cite{elli2}. Particularly, from the point of view of $q$-deformed $SU(2)$ 
algebra, we can adopt the  representation 
which is manifestly free of negative norm
\ba
a_q &=& \sqrt{[1-n_0]_q+[n_0]_q}|0\rangle\langle 1| + \cdots
+ \sqrt{[1-n_0]_q+[n_0]_q}|s-1\rangle\langle s|,\\
a^{\dag}_{q} &=& (a_q)^{\dag},\\
N^{\prime} & \equiv & N-n_0 = -n_0|0\rangle\langle 0| +
 (1-n_0)|1\rangle\langle 1|+
\cdots + (s-n_0)|s\rangle\langle s|,
\ea
and $n_0 = {(s+1)\over 4}$, \cite{fuji}.
It is  easy to see that deformed annihilation
and creation operators
$a_q$ and $a^{\dag}_{q}$  
also satisfy the identical linear relations,
 Eqs. (\ref{os1}) and (\ref{os2})
respectively,
 and they
can be derived from the dynamics of unitary phase operator Eq. (\ref{dyo})
and polar decomposition, $a_q = e^{i\Phi}\sqrt{[N-n_0]_q+[n_0]_q}$ .

Now consider two such type of commuting $q$-deformed oscillators,
$\{a_q,a^{\dag}_{q},N_1\}$ and $\{b_q,b^{\dag}_{q},N_2\}$. 
They can be used to realize generators of $q$-deformed
$SU(2)$ algebra \cite{fuji}, by the generalized Jordan-Schwinger mapping 
\cite{9b} 
\be
\tilde{J}_+ = a^{\dag}_{q}b_q, \quad \tilde{J}_-=b^{\dag}_{q}a_q,
 \quad J_0 =(N_1-N_2)/2.
\label{js}
\ee 
Then taking the hamiltonian of a two-dimensional harmonic 
oscillator $H = \hbar(N_1 \omega_1 + N_2 \omega_2)$,
we again see  that the dynamics   
 for $\tilde{J}_{\pm,0}$ is identical to that of  ${J}_{\pm,0}$ operators 
 {\it i.e.} Eqs. (\ref{eq2a}) and (\ref{eq2b}),
if we impose the condition $\omega_2-\omega_1  =\mu B$. 
In this sense, under the mapping in Eq. (\ref{js}), 
The dynamics of $\tilde{J}_{\pm,0}$
follows from the dynamics of unitary phase operator for a 
finite dimensional $q$-deformed oscillator.

Concluding, the Wigner problem for a precessing
magnetic dipole has been analyzed  through the polar 
decomposition of ladder operators. We have argued that 
a dynamics in terms of unitary phase operator  can be assigned to   
various types of deformations of ladder operators.
\section*{Acknowledgements}
The author would like to acknowledge Alexander von Humboldt Foundation, 
Germany for grant of financial support. 

\end{document}